\documentstyle[11pt,aaspp4]{article}
\tighten

\def\check_mode#1{\ifmmode{#1}\else{$#1$}\fi}
\def\deg    {\check_mode{^\circ}}
\def\ddeg   {\check_mode{{\rlap.}^\circ}}
\def\lsim   {\check_mode{_<\atop^{\sim}}}
\def\um     {$\mu$m}
\def\icm    {cm$^{-1}$}

\begin{document}

\def\calpaper{Fixsen \etal\ 1994b}
\hfuzz=10pt \overfullrule=0pt

\def\blankline{\par\vskip \baselineskip}

\def\Received{\par\vskip 36 pt
        \centerline{\it Received  \vrule height -1 pt width 2.5 in
        depth 1.8 pt .}\par\blankline}

\def\refitem #1! #2! #3! #4;{\hang\noindent
    \hangindent 20pt\rm #1, \it #2, \bf #3, \rm #4.\par}
\def\bookref{\par\noindent\hangindent 20pt}

\def\folio{\ifnum\pageno=1\nopagenumbers\else\number\pageno\fi}

\def\wisk#1{\ifmmode{#1}\else{$#1$}\fi}

\def\lt     {\wisk{<}}
\def\gt     {\wisk{>}}
\def\le     {\wisk{_<\atop^=}}
\def\ge     {\wisk{_>\atop^=}}
\def\lsim   {\wisk{_<\atop^{\sim}}}
\def\gsim   {\wisk{_>\atop^{\sim}}}
\def\kms    {\wisk{{\rm ~km~s^{-1}}}}
\def\Lsun   {\wisk{{\rm L_\odot}}}
\def\Msun   {\wisk{{\rm M_\odot}}}
\def\um     {\wisk{{\rm \mu m\ }}}
\def\sig    {\wisk{\sigma}}
\def\etal   {{\sl et~al.\ }}
\def\bsl    {\wisk{\backslash}}
\def\by     {\wisk{\times}}

\def\amin   {\wisk{^\prime\ }}
\def\asec   {\wisk{^{\prime\prime}\ }}
\def\cc     {\wisk{{\rm cm^{-3}\ }}}
\def\deg    {\wisk{^\circ\ }}
\def\ddeg   {\wisk{{\rlap.}^\circ}}
\def\damin  {\wisk{{\rlap.}^\prime}}
\def\dasec  {\wisk{{\rlap.}^{\prime\prime}}}
\def\approxeq{$\sim \over =$}
\def\abouteq{$\sim \over -$}
\def\percm{cm$^{-1}$}
\def\percmsq{cm$^{-2}$}
\def\percmcub{cm$^{-3}$}
\def\perhz{Hz$^{-1}$}
\def\perpc{$\rm pc^{-1}$}
\def\persec{s$^{-1}$}
\def\peryr{yr$^{-1}$}
\def\te{$\rm T_e$}
\def\tenup#1{10$^{#1}$}
\def\to{\wisk{\rightarrow}}
\def\thin{\thinspace}
\pretolerance=1000	
\raggedright
\title{Comparison of the {\it COBE}\altaffilmark{1} FIRAS and DIRBE Calibrations}

\author{ D.J. Fixsen\altaffilmark{2,6},
         J.L. Weiland\altaffilmark{2},
         S. Brodd\altaffilmark{2},
         M.G. Hauser\altaffilmark{3},
         T.   Kelsall\altaffilmark{4},
         D.T. Leisawitz\altaffilmark{5},
         J.C. Mather\altaffilmark{4},
	 K.A. Jensen\altaffilmark{2},
         R.A. Shafer\altaffilmark{4},
         R.F. Silverberg\altaffilmark{4}}

\altaffiltext{1}{The National Aeronautics and Space Administration/Goddard 
Space Flight Center (NASA/GSFC) is responsible for the design, development, 
and operation of the Cosmic Background Explorer ({\it COBE}).
Scientific guidance is provided by the {\it COBE} Science Working Group.
GSFC is also responsible for the development of the analysis software and
for the production of the mission data sets.}
\altaffiltext{2}{Hughes STX Corporation, 
                 Code 685.9, NASA/GSFC, 
                 Greenbelt MD 20771.}
\altaffiltext{3}{Space Telescope Science Institute,
                 3700 San Martin Drive,
                 Baltimore, MD 21218.}
\altaffiltext{4}{Code 685, NASA/GSFC,
                 Greenbelt MD 20771.}
\altaffiltext{5}{Code 631, NASA/GSFC,
                 Greenbelt MD 20771.}
\altaffiltext{6}{e-mail: fixsen@stars.gsfc.nasa.gov}

\begin{abstract}

We compare the independent FIRAS and DIRBE observations from the {\it COBE}
in the wavelength range 100-300\um. This cross calibration provides checks of 
both data sets.  The results show that the data sets are consistent within
the estimated gain and offset uncertainties of the two instruments. They
show the possibility of improving the gain and offset determination
of DIRBE at 140 and 240 \um.
\end{abstract}


\section{Introduction} 

The measurement of the sky brightness with the Far InfraRed Absolute Spectrophotometer
(FIRAS) and the Diffuse InfraRed Background Experiment (DIRBE)  
instruments (Boggess \etal 1992; Mather \etal 1993; Silverberg \etal 1993) 
provides important information on the infrared sky.  
The two instruments
were both designed to determine all-sky absolute brightness, but 
calibrations of these instruments were done in different and independent ways.
FIRAS is a spectrometer, and DIRBE a broad-band photometer; their
measurements overlap in the wavelength range 100 - 300 \um.
A natural question is: ``do these two
very different instruments find the same results where their measurements
overlap?"  Comparison of the two
provides a check on the accuracy of the sky measurements.

The FIRAS utilizes a polarizing Fourier-transform spectrometer to measure the 
difference in radiation from two inputs.  One input is connected to an internal
blackbody, the other is connected to either the sky or an external
blackbody calibrator. The FIRAS derives its absolute calibration from the 
temperature controlled external blackbody that completely fills the input sky 
horn. In nine calibration events (approximately one each month) over the 10 
month mission the external blackbody was placed into the mouth of the horn.  
Then over the course of several days the temperature was varied from 2.2K to 
20K. There were also eight calibrations where the external calibrator was 
kept cold. For the cold data (T$_x\sim$2.7 K), the radiation was essentially
zero at frequencies 30 to 100 \icm (300 to 100 \um) and so allows determination
of instrumental offsets. The high temperature data (T$_x\sim$20 K) 
then allow determination of the gain.  The systematic offset uncertainties 
of the FIRAS can be calculated from the ``PEP" uncertainties
(FIRAS Explanatory Supplement; Fixsen \etal\ 1994b). 
The systematic gain uncertainties from the calibration model itself are 
$\sim$0.2\%. However, there are inaccuracies in the bolometer model
that lead to larger errors.  Comparing the various channels (four detectors
and two modes of operation) leads to estimates of $\sim$0.5\% for the gain
uncertainty.  As there is the possibility that the error in the bolometer model
could bias all of the detectors in the same direction, we use a conservative
estimate of the uncertainty of 2\% for the high frequency FIRAS data 
(20-80 \icm). At the highest FIRAS frequencies (80-100 \icm), the uncertainty
on the temperature of the external blackbody during calibration increases 
the gain uncertainty exponentially with frequency as this
is on the Wein side of the Planck function. The gain has
been verified at lower frequencies to 0.5\% by comparison with data from the
COBE/DMR experiment (Fixsen \etal\ 1994a).

The DIRBE is an absolute radiometer, which measures sky brightness by
chopping between the sky and a zero-flux internal reference at 32 Hz.
Instrumental offsets were measured roughly five times per orbit by closing
a cold shutter located at the prime focus.  A constant radiative offset signal
in the long-wavelength detectors was found to originate from the JFET 
amplifiers operating at about 55 K within the dewar.
This offset was measured via the shutter-closed calibrations and
removed from the data; systematic uncertainty in the offset correction
is approximately 160, 250 and 30 kJy/sr at 240, 140, and 100 \um\ 
respectively (DIRBE Explanatory Supplement, 1997).
Relative gain stability of $\sim$1\% was achieved over the course
of the mission by frequently monitoring internal radiative reference sources
and isolated bright celestial sources.  Measurements were placed on an
absolute photometric scale through the use of Voyager observations and 
atmospheric models of Jupiter (140 and 240 \um) and Uranus (100 \um).  The 
uncertainty in the absolute gain at these wavelengths is $\sim$10\% 
(DIRBE Explanatory Supplement, 1997).

\section{The Data}

FIRAS and DIRBE data are pixelized into maps using a common projection and 
coordinate system (geocentric ecliptic coordinates, epoch 2000).
The projection, referred to as the quadrilateralized sphere
(O'Neill \& Laubscher, 1976; DIRBE Explanatory Supplement, 1997), 
approximately preserves
area.  The FIRAS and DIRBE pixels are related such that there are
64 DIRBE pixels for every FIRAS pixel.  Full FIRAS coverage of the
sky is given in 6144 pixels; DIRBE has 393216 pixels.

The high frequency FIRAS ``Pass 4" data consist of 167 point spectra between 
20 and 100 cm$^{-1}$ (500 $\mu$m to 100 $\mu$m) in each of 6067 pixels on 
the sky (77 pixels have no data).  They were calibrated using the 
method described in \calpaper, with the improvements noted in Fixsen \etal\ 
1996. A weighted average of all of the high frequency FIRAS data was used.
The weights and uncertainties are discussed in \calpaper. 

For the DIRBE counterpart to the FIRAS data,
we used the DIRBE 240 \um, 140 \um and 100 \um\ (bands 10, 9  and 8) 
``Annual Average Maps" from the 1997 Pass 3b production software release.
``Annual Average Maps" are weighted averages of the first 10 months of 
data taken during the cryogenic period (11 Dec 1989 to 21 Sep 1990); these
maps are described in the DIRBE Explanatory Supplement (1997).  
Random uncertainties associated with the DIRBE measurements are also taken from
the Annual Average product; these are the standard deviations of the
photometric mean reported for each pixel. The use of the Annual 
Average Maps is essential for the band 10 and 9 comparisons, since
many observations are needed to achieve the best DIRBE signal-to-noise ratio
possible at these wavelengths in dark parts of the sky. However,
since FIRAS sky measurements are all made at a solar elongation, $e$, 
of 94\deg, there is an error introduced by comparing the FIRAS dataset 
with the DIRBE Annual Average Maps, which sample a mix of solar elongation 
angles for each pixel.  The error is wavelength dependent and diminishes with
increasing wavelength, as it is the emission from the
interplanetary dust cloud which introduces the discrepancy.  
At 100 $\mu$m, the 1-$\sigma$ photometric error associated with using the 
Annual Average Map rather than a map interpolated to $e = 94$\deg is 
$\sim 3$\% (of order 100 kJy/sr at the Galactic poles).  
Darker regions ($|b| \gt 30$\deg), where 
zodiacal emission can contribute up to 60-70\% of the observed sky brightness,
are more affected than the Galactic plane.  As will be seen in section 3,
the use of the Annual Average band 8 map, while not ideal, is not
the limiting uncertainty in the comparison with the FIRAS data.
At 140 and 240 \um, emission from the Galaxy dominates over that from the 
interplanetary dust at nearly all latitudes, and the band 9 error associated 
with using the Annual Average is estimated as $\lt 1$\%  ($\sim 40$ kJy/sr)
at high Galactic latitudes, and completely negligible for all parts of the sky
in band 10.

\section{Data Preparation}

Before comparing the FIRAS and DIRBE maps, the data must be converted to
a common spectral and spatial format.  The spectral information of the FIRAS
is converted into a broadband photometric measurement like the
DIRBE's, and the DIRBE spatial resolution is degraded to that of the FIRAS.

\subsection{FIRAS}

For each 3\deg FIRAS pixel, the high frequency spectral data,
$I_F(\nu)$, are convolved with 
the DIRBE system responsivity, $R_D(\nu)$ (see fig 1), in order to produce 
two maps to compare to DIRBE bands 9 and 10.
Since DIRBE flux densities are quoted in each passband using an assumed 
intrinsic spectral energy distribution $\nu I_{\nu} = constant$,
we use the following equation to compute the corresponding FIRAS 
brightness in each pixel:
\begin{equation}
I_F^{\nu_0} = \frac{\int I_F(\nu) R_D(\nu) d\nu}
{\int \frac{\nu_0}{\nu} R_D(\nu) d\nu}
\end{equation}
where $\nu_0$ is the frequency corresponding to the DIRBE nominal
wavelength, and integration covers the passband.

Special treatment is required for the DIRBE band 8, as the FIRAS 
spectral data do not extend to the short-wavelength edge of the DIRBE passband.
We construct an alternate truncated passband over which the FIRAS data may be 
integrated, and then use a ``color correction" factor to derive the 
equivalent value for the true DIRBE passband. 
We consider two truncated passbands, $R_{8x}(\nu)$.
The first, $R_{8'}(\nu)$, uses a spectrum tuned 
to the FIRAS noise, and has a lower mean frequency (and hence a larger color 
correction). The second passband, $R_{8''}(\nu)$, is a better match to the 
DIRBE band 8, but weights frequencies for which the FIRAS data
have higher intrinsic noise.
Both $R_{8'}(\nu)$ and $R_{8''}(\nu)$ are plotted in figure 1 to
show the relationship to the actual DIRBE 100 $\mu$m bandpass.

The calculation of the color correction is dependent upon the contributors
to the sky flux in the bandpass.  Since both zodiacal and Galactic emission
are important contributors at wavelengths near 100 $\mu$m, we rewrite the 
observed FIRAS sky brightness $I_{F}(\nu)$ as the sum of ``zodiacal" and
``non-zodiacal" terms: 
\begin{equation}
I_{F}(\nu) = Z(\nu) + [I_{F}(\nu) - Z(\nu)], 
\end{equation}
where the term in square brackets is expected to be dominated by Galactic 
emission.  The FIRAS band 8 map is then computed using
\begin{equation}
I_F^{\nu_0} = C_{8x}^{Z} \frac{\int [Z(\nu)] R_{8x}(\nu)
d\nu}{\int \frac{\nu_0}{\nu} R_{8x}(\nu) d\nu} +
C_{8x}^{I-Z} \frac{\int [I_F(\nu)-Z(\nu)] 
R_{8x}(\nu) d\nu}{\int \frac{\nu_0}{\nu} R_{8x}(\nu) d\nu} 
\end{equation}
where $C_{8x}^{Z}$ and $C_{8x}^{I-Z}$ are color corrections appropriate
to each component.  We use the DIRBE interplanetary dust model
(Reach \etal 1996; Kelsall \etal 1997) to evaluate the zodiacal 
contribution $Z(\nu)$.

To compute the color corrections $C_{8x}$, we assume that each of the two 
sky components may be approximated with a spectrum of form
$\alpha \nu^\beta B(T,\nu)$ where $\alpha,~~\beta$ and $T$ are parameters 
of the fit and $B(T,\nu)$ is the Planck function.  The color correction is 
then estimated as:
\begin{equation}
C_{8x} = \frac{\int \nu^\beta B(T,\nu) R_D(\nu) d\nu}
{\int \nu^\beta B(T,\nu) R_{8x}(\nu) d\nu}
\end{equation}

To determine the spectral model for the $(I-Z)$ component we subtracted the 
DIRBE model of the 
zodiacal emission, $Z$, from the FIRAS emission, then estimated the temperature 
and index of the remaining average spectrum over the entire sky (G) or only over
the $|b|>20^\circ$ region (H). 
The parameters of the spectrum fits are: 
($\beta, T$)=(1.72, 19.9),(1.54, 19.6) and  
(.6, 116) for the full sky (G), high latitude sky (H) and the 
zodiacal model respectively.
These give color corrections of .78, .74 and 2.19 for the $8'$ passband and
.83, .81 and 1.5 for the $8''$ passband.  
The spectral variation of different sources gives rise to a range of color 
corrections. The color corrections given here are reasonable, but variations 
of $\sim10$\% over the sky or under other assumptions are possible.

\subsection{DIRBE}

The FIRAS data have $\sim 7^{\circ}$ angular resolution, as measured 
by their full width at half maximum (FWHM), but the DIRBE beam profiles are 
$\sim 0.7^{\circ}$. To compare these two data sets, the DIRBE data were 
convolved with a map of the FIRAS beam obtained from on orbit data taken when
the moon was near the FIRAS beam (See Fig 2).
The COBE rotates about the optical axis of the FIRAS instrument, so 
{\it on average} the beam must have cylindrical symmetry.  
However, the time it takes to collect a single interferogram is less than 
a rotation period, so a
particular measurement beam may be asymmetric. The dwell time for
each FIRAS interferogram also causes the beam to be elongated in the scan 
direction by 2.4\deg. To account for this, the DIRBE data were further
convolved one FIRAS pixel (2.6\deg) along ecliptic meridians (close to the
FIRAS scan direction). The convolved DIRBE map data were then 
decimated to obtain three 6144 pixel all sky maps at the same resolution 
as the FIRAS data. We estimate that the assumption of beam symmetry
may produce residual beam shape errors of order 5\%. These errors are only 
important in the high gradient regions near the Galactic plane.

The mean position of the FIRAS data is not centered within
each pixel.  While this is a minor matter in the high Galactic latitude regions,
near the plane it has a significant effect.  To correct for this effect
we fit a general second order polynomial (six parameters) to each pixel and its 
eight closest neighbors on the DIRBE maps after beam convolution and
reduction to FIRAS resolution.  The difference between the value
of this polynomial at the mean location of the FIRAS data and the value at the 
pixel center is the ``gradient correction". The ``gradient correction" was
applied to the DIRBE data for each pixel. A typical correction (rms) is 5\%.

\section{Comparison}

Figure 3 presents plots of the FIRAS vs. DIRBE data 
after the data conversion steps described above. There is a
clear correlation between the measurements
in all 3 bands.  The noise appears higher in the FIRAS data than in the 
DIRBE data:  part of this is due to the larger variation in the FIRAS sky
coverage, which leaves some pixels with very few observations and hence large
errors. 

We make a formal fit 
\begin{equation}
I_F = g I_D + o,
\end{equation}
with free gain $g$ and offset $o$ parameters, to fit the FIRAS data to the 
DIRBE data for each DIRBE band. 
Our first set of fits allowed only for the random errors in
the DIRBE and FIRAS data; the results for these are presented in Table~1
and are indicated with an asterisk.
The results are encouraging but the formal $\chi^2$ is large: a problem 
is evident in regions of high gradient. 

The beam shape and the gradient correction errors might account for the 
large $\chi^2$, but the residual DIRBE gain uncertainties and residual 
stripes in the FIRAS data might also be significant.
To account for these, we include in the uncertainty estimate 5\% for the
beam error and a 0.5$^\circ$ gradient correction error. 
Both sets of statistics are
included in Table 1 to show how these affect the gain and offset determination
between the instruments. We include fits over the entire sky, which are 
listed in Table 1 as the results for a Galactic cut of 0\deg.
Alternatively, we can restrict our comparison to the high Galactic 
latitude regions ($|b| \gt 10^\circ~~or~~20^\circ$).  This reduces the 
available dynamic range over which to compare the gains, but also acts to
reduce the biases in the derived offset.

Quadratic fits of the form $I_F = q I_D^2 + g I_D + o$
are included to test the linearity of the instruments.
Since the brightest part of the sky is $\sim 10^6$ kJy/sr, the fifth column
in Table 1 gives an estimate of the fraction of the nonlinearity over the 
full range. The Galactic plane is needed to give a good estimate for the
quadratic parameter.

\section{Discussion}

There are several issues that are apparent from the results listed in Table 1.
The statistical uncertainties listed there are clearly not the major uncertainty
in comparing the two data sets.  The difference in the fits with and without
the Galactic plane is consistently a larger factor. This might be a problem
with detailed pointing, beam shape problems, color variation differences coupled
with errors in the response function, or variation in zodiacal emission.
All of these problems are larger at higher frequency and many of them are
larger in the Galactic plane. 

Even with these limitations, for DIRBE bands 10 and 9, there is no evidence
for non-linearity: the gain ratio can be determined to within a 
percent or so, and the offset can be roughly determined.  
For the gain ratio and quadratic term,
the estimates including the Galactic plane are better, because they have a much
longer lever arm, as indicated by the statistical uncertainties. Still, the
variation when the Galactic plane is excluded should not be ignored, but used
as evidence of the real uncertainty of the comparison.  To find the offset 
between the measurements, removing the Galactic plane eliminates many of the 
problems and uncertainties of the comparison as indicated by the $\chi^2$ but
leaves 2/3 of the sky to compare.  Again, the differences between the other 
fits are an indicator of the real uncertainty.

The comparison between FIRAS and DIRBE band 8 has all of the caveats of the 
comparison to bands 9 and 10, plus those associated with the ``color" 
correction.
Many other schemes for comparing the FIRAS data with the DIRBE band 8
data are possible; the methods shown here are only meant to convey the range 
of solutions that result from reasonable sorts of fits.  
We note that there is an inconsistency in tabulating all-sky fits for
the high-latitude (H) solutions, and similarly in showing fits to data cut at 
$b=20\deg$ for the all-sky (G) solutions.  However, we include these within 
Table~1 to show that the dominant effect is that of latitude cut rather than 
color correction.  We adopt a similar philosophy for the entries in
which the data have been cut at $b=10\deg$, but color corrected to match
either the G or H solutions.
The high $\chi^2$ obtained for the 100 $\mu$m fits is a 
reminder that there are still problems.  The extrapolation has the intrinsic
problem that any emission beyond 97 \icm\ cannot be observed by the FIRAS
instrument.

In addition to the random uncertainties and the uncertainties generated in the 
process to obtain comparable data, there are the systematic uncertainties 
associated with each instrument. While the random uncertainties are lower
in the DIRBE data, the FIRAS calibration is done directly to a 
temperature-controlled blackbody source, and this provides the FIRAS with lower
absolute uncertainties in gain and offset. 

\section{Conclusions}

After examining the fits in Table 1 and other permutations on fitting options,
we note the uncertainties are not dominated by the statistical uncertainties
but by other effects.  We have adopted values for the gain and offset
based upon results from those fits in Table~1 which are best suited for 
determining these parameters.
For band 8, this means the $8''$ bandpass is probably
better because it is a closer frequency match.
The offsets were selected from the fits excluding the galactic plane, including
comparison uncertainties. The gains were chosen from fits including the 
Galactic plane and
comparison uncertainties. The final results, uncertainties and upper limits
for a quadratic effect are summarized in Table 2. The uncertainties are 
the dispersion in the results from Table 1. They are dominated by systematic
effects in the comparison rather than the intrinsic noise, but the results
still test the DIRBE calibration.

The FIRAS spectra convolved with the DIRBE pass bands agrees with the DIRBE 
observations convolved with the FIRAS beam within the estimated DIRBE 
uncertainties.  This lends support to the stated DIRBE uncertainties in both 
the gain and offset. The comparison uncertainties provide a poor check on DIRBE 
band 8. However, the lower FIRAS systematic uncertainties could be used to 
recalibrate DIRBE bands 9 and 10 to reduce their systematic uncertainties. 

This work was supported by {\it COBE} Extended Mission funding from the NASA
Office of Space Sciences.

\clearpage
\begin{table}[h]
\label{cc_stat_table1}
\caption{FIRAS-DIRBE Cross-Calibration Results}
\begin{centering}
\begin{tabular}{ccrccr}\hline \hline

 Band & Galaxy Cut & Offset & Gain & Quadratic & $\chi^2$/DOF \\
      & (degrees) & (kJy/sr) & - & ($10^{-6}$sr/kJy) &-~~~~ \\ \hline
10$^*$ & 0 & $-177\pm  3$ & $1.0600\pm 0.0001$ & - &            472.09 \\
10$^*$ &20 & $-133\pm  5$ & $1.0462\pm 0.0007$ & - &              3.29 \\
10 & 0 & $-185\pm  6$ & $1.0566\pm 0.0012$ & - &              1.04 \\
10 &10 & $-165\pm  7$ & $1.0511\pm 0.0015$ & - &              1.21 \\
10 &20 & $-137\pm  8$ & $1.0425\pm 0.0019$ & - &              1.03 \\
10 & 0 & $-177\pm  6$ & $1.0546\pm 0.0013$ & $0.04\pm 0.01$ & 1.04 \\
\hline
9$^*$ &  0 & $-917\pm 15$ & $1.0539\pm 0.0002$ & - &             43.19 \\
9$^*$ & 20 & $-509\pm 30$ & $1.0183\pm 0.0031$ & - &              1.61 \\
9 &  0 & $-671\pm 24$ & $1.0408\pm 0.0018$ & - &              1.39 \\
9 & 10 & $-627\pm 29$ & $1.0344\pm 0.0028$ & - &              1.70 \\
9 & 20 & $-515\pm 36$ & $1.0174\pm 0.0043$ & - &              1.39 \\
9 &  0 & $-647\pm 26$ & $1.0379\pm 0.0022$ & $0.02\pm 0.01$ & 1.39 \\
\hline
${8'}_G^*$ &  0 & $-1596\pm  36$ & $1.2414\pm 0.0005$ & - &          11.55 \\
${8'}_G^*$ & 20 & $ -196\pm  56$ & $1.0416\pm 0.0068$ & - &           1.58 \\
$8'_G$ &  0 & $-1483\pm  35$ & $1.2381\pm 0.0028$ & - &              1.89 \\
$8'_G$ & 10 & $ -834\pm  51$ & $1.1463\pm 0.0055$ & - &              2.00 \\
$8'_G$ & 20 & $ -305\pm  61$ & $1.0564\pm 0.0077$ & - &              1.51 \\
$8'_G$ &  0 & $-1465\pm  40$ & $1.2358\pm 0.0037$ & $0.02\pm 0.02$ & 1.89 \\
${8'}_H^*$ &  0 & $-1323\pm  22$ & $1.1779\pm 0.0005$ & - &          11.50 \\
${8'}_H^*$ & 20 & $ -123\pm  54$ & $1.0066\pm 0.0065$ & - &           1.57 \\
$8'_H$ &  0 & $-1244\pm  34$ & $1.1777\pm 0.0028$ & - &              1.82 \\
$8'_H$ & 10 & $ -703\pm  48$ & $1.1006\pm 0.0053$ & - &              1.95 \\
$8'_H$ & 20 & $ -255\pm  58$ & $1.0251\pm 0.0073$ & - &              1.49 \\
$8'_H$ &  0 & $-1239\pm  38$ & $1.1771\pm 0.0036$ & $0.00\pm 0.02$ & 1.82 \\
\hline
${8''}_G^*$ & 0 & $-2693\pm 124$ & $1.3100\pm 0.0028$ & - &           3.07 \\
${8''}_G^*$ &20 & $   54\pm 299$ & $1.0210\pm 0.0361$ & - &           2.46 \\
$8''_G$ & 0 & $-1801\pm 134$ & $1.2495\pm 0.0054$ & - &              2.59 \\
$8''_G$ &10 & $ -265\pm 243$ & $1.0909\pm 0.0248$ & - &              3.11 \\
$8''_G$ &20 & $   74\pm 302$ & $1.0181\pm 0.0365$ & - &              2.45 \\
$8''_G$ & 0 & $ -557\pm 158$ & $1.1289\pm 0.0098$ & $0.40\pm 0.03$ & 2.55 \\
${8''}_H^*$ & 0 & $-2542\pm 121$ & $1.2785\pm 0.0027$ & - &           3.07 \\
${8''}_H^*$ &20 & $   80\pm 292$ & $1.0046\pm 0.0352$ & - &           2.46 \\
$8''_H$ & 0 & $-1660\pm 131$ & $1.2187\pm 0.0054$ & - &              2.58 \\
$8''_H$ &10 & $ -208\pm 238$ & $1.0690\pm 0.0243$ & - &              3.11 \\
$8''_H$ &20 & $  100\pm 294$ & $1.0019\pm 0.0357$ & - &              2.45 \\
$8''_H$ & 0 & $ -464\pm 155$ & $1.1024\pm 0.0096$ & $0.39\pm 0.03$ & 2.54 \\
\hline
\end{tabular}
\end{centering}

{\small $^*$ These fits use the random uncertainties 
of the FIRAS and DIRBE data only.}

{\small The $8'$ and $8''$ passbands use 
different extrapolations of the FIRAS data to compare to the DIRBE data.}

{\small $G H$ The $8_G$ and $8_H$ use different color 
corrections of the FIRAS data to compare to the DIRBE data. See text.}

\end{table}

\begin{deluxetable}{crrl}
\tablewidth{5.5in}
\tablecaption{Adopted FIRAS-DIRBE Calibration}
\tablehead{ \colhead{DIRBE} &
            \colhead{Offset}   &
            \colhead{Gain} &
            \colhead{Quadratic limits} \nl
            \colhead{Band}  &
            \colhead{kJy/sr}   &
            \colhead{-}   &
            \colhead{$10^{-6}$sr/kJy} }
\startdata
10 & $-137\pm  22\pm  20$ & $1.06\pm 0.01\pm 0.02$ & 0.05 \nl
9  & $-515\pm 150\pm 120$ & $1.04\pm 0.02\pm 0.02$ & 0.04 \nl
8  & $100\pm 830\pm 900$ & $1.25\pm 0.10\pm 0.15$ & 0.2  \nl
\enddata
\tablenotetext{}{Uncertainties are dominated by systematic effects.
Second uncertainties are the absolute uncertainties of the FIRAS data.}
\label{cc_stat_table2}
\end{deluxetable}

\clearpage

\figcaption[]{DIRBE passbands and FIRAS approximations to DIRBE band 8.
The DIRBE 10, 9 and 8 bands are plotted as solid lines.
The $8'$ and $8''$ band constructs are plotted as dotted
and dot-dash lines, respectively. The $8'$ and $8''$ approximations
are necessary because the FIRAS data ends at 97 \icm ; the truncation
has an insignificant effect on band 9. Color corrections
must be applied to band 8 results.
\label{fspectra}
}

\figcaption[]{FIRAS Beam response as function of angle from central axis.
\label{beam}
}

\figcaption[]{Scatter plots of the DIRBE data vs the FIRAS data, after
conversion to a common format.  The solid line corresponds to the adopted
fit given in Table~2.
a) 240 $\mu$m (Band 10)  b) 140 $\mu$m (Band 9)  c) 100 $\mu$m
(Band 8, using the $8''$ bandpass). 
\label{fcuts}
}

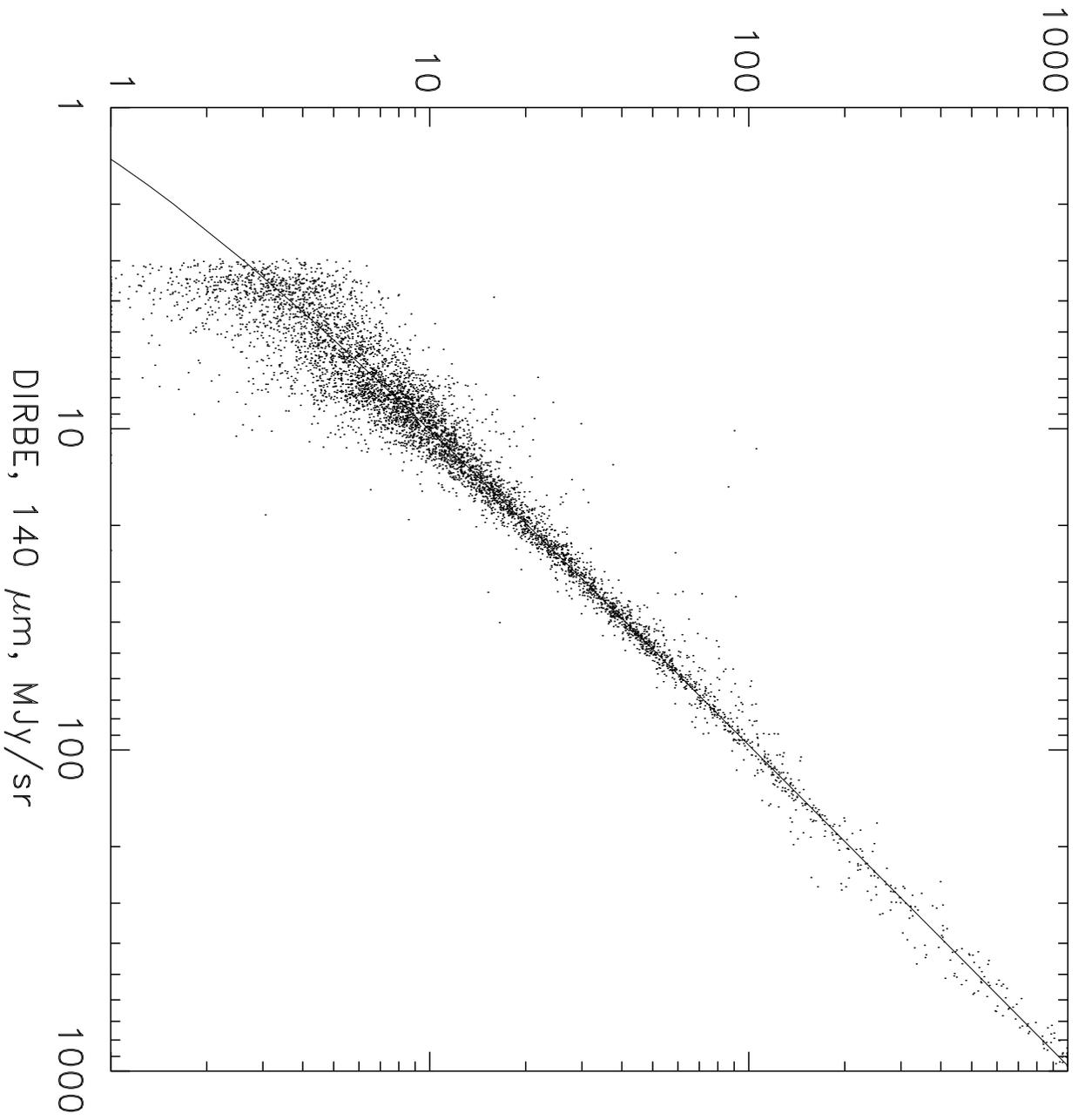
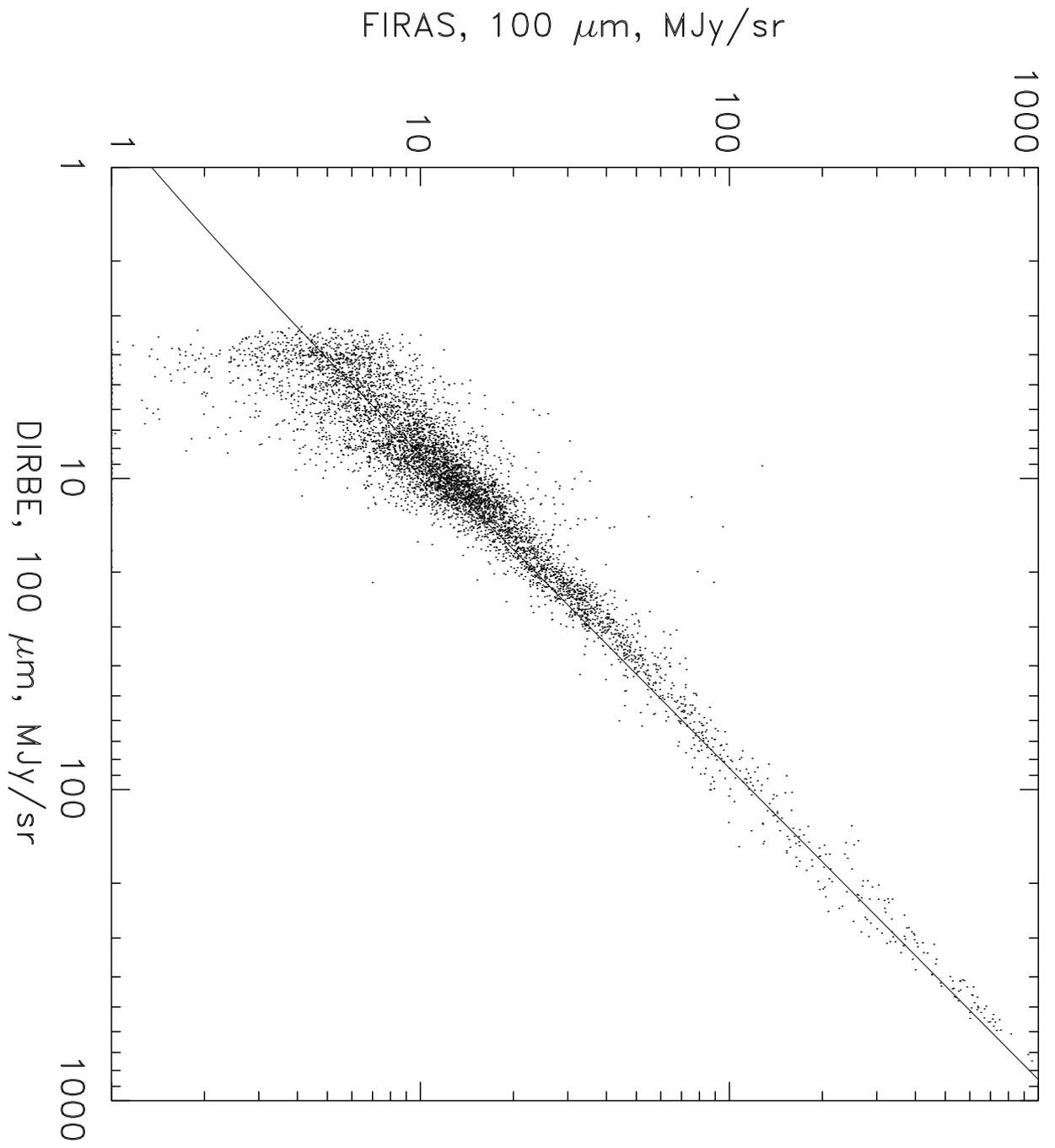
\end{document}